\documentclass{aa}  

\usepackage{graphicx}
\usepackage{txfonts}

\usepackage{amsmath}	
\usepackage{amssymb}	
\usepackage{xspace}
\usepackage{gensymb}
\usepackage{pdflscape}
\usepackage{xcolor}
\usepackage[export]{adjustbox}
\usepackage{placeins}

\usepackage{natbib,twoopt}
\usepackage[breaklinks=true]{hyperref} 
\bibpunct{(}{)}{;}{a}{}{,}             

\newcommand{\ter}{Terzan~5\xspace}
\newcommand{\ergs}{erg~s$^{-1}$\xspace}
\newcommand{\chandra}{{\it Chandra}\xspace}

\newcommand{\rh}{$R_h$\xspace}
\newcommand{\nh}{$N_{\rm H}$\xspace}

\begin{document}

   \title{A deep \chandra study verifies diffuse non-thermal X-ray emission from the globular cluster Terzan 5}


   \author{Jiaqi Zhao
          \inst{1} 
          \and
          Craig O. Heinke\inst{1}
          \and
          Su Fu\inst{1}
          }

   \institute{Physics Department, CCIS 4-183, University of Alberta, Edmonton, AB, T6G 2E1, Canada\\
              \email{jzhao11@ualberta.ca; heinke@ualberta.ca}
             }

   \date{Received xxx; accepted yyy}

 
  \abstract
   {Diffuse X-ray emission has been detected from a few Galactic globular clusters (GCs), whereas its nature still remains largely unclear. The GC Terzan~5 was previously found 
   to show 
   a significant diffuse thermal X-ray excess from its field, likely contributed by the 
   Galactic background, and a non-thermal component described by a power-law model with photon index $\Gamma \sim 1$.}
   {
   With over 16 times the accumulated \chandra\ exposure time as in the prior study, 
   we are motivated to reexamine and verify the diffuse X-ray emission from the field of \ter, 
   enabling constraints on its nature. 
   }
   {We analyze all the available and useful \chandra observations of \ter, including 18 observations over a span of 13 years, with a total exposure time of 641.6 ks. To study the diffuse X-ray emission, we focus on {four annular regions with an equal width of 0.72~arcmin} centered on \ter (0.72--3.60 arcmin), where we extract and analyze the X-ray spectra, after removing point sources and instrumental backgrounds.}
   {We verify a significant diffuse X-ray excess from the field of \ter in the band 0.8--3 keV.
   {After constraining the contribution from local X-ray background, we find a diffuse X-ray component that is genuinely associated with \ter, which can be well described by a power-law model. More interestingly, the fitted photon indices show a significant increase from $\Gamma = 1.96 \pm 0.18$ in the inner region to $\Gamma = 3.48 \pm 0.71$ in the outer region. The diffuse X-rays can also be well fitted by a thermal bremsstrahlung model, with plasma temperatures declining from $kT \sim 3$~keV to $kT \sim 1$~keV. }
   }
   {We suggest that {synchrotron radiation from the combined pulsar winds of \ter's millisecond pulsar population is a possible origin of the observed diffuse X-ray emission, 
   but 
   the the large steepening in the spectra cannot be produced solely by synchrotron cooling. Other radiation processes, like thermal bremsstrahlung, may also contribute to the diffuse X-rays.}}


   \maketitle
%

\section{Introduction}

Globular clusters (GCs) are dense, gravitationally bound stellar systems composed of an old stellar population.
As a consequence of high stellar densities and interaction rates, GCs are found to 
dynamically produce some kinds of  
X-ray sources \citep[see e.g.][]{Pooley03,Heinke03d,Bahramian13}, especially low-mass X-ray binaries (LMXBs) with neutron stars accreting from companion stars. 
Some 22 LMXBs in Galactic globular clusters have been seen in a bright state, $L_X>10^{35}$ \ergs \citep{BahramianDegenaar23}, most of which show transient bright outbursts of order a month. There is also a much larger population of quiescent LMXBs, $L_X<10^{34}$ \ergs, that have not yet been seen 
{in} outburst \citep{Heinke03d,Guillot09}.  
Other X-ray sources in globular clusters include cataclysmic variables (CVs) including an accreting white dwarf \citep{Edmonds03a}, chromospherically active binaries (ABs) composed of normal stars \citep{Dempsey97,Grindlay01a,Bassa04,Heinke2005}, and millisecond pulsars (MSPs).  
MSPs are descendants of LMXBs, and like them are typically produced dynamically in dense clusters \citep{Hui11,Bahramian13}.  
MSPs are typically identified via their radio pulsations \citep[e.g.][]{Lorimer08}. However, their substantial non-thermal gamma-ray emission has also been detected in several globular clusters \citep[e.g.][]{Abdo2009,Abdo10}, including \ter \citep{Kong10}, which has the highest known population of MSPs of any cluster (49; \citealt{Padmanabh24}).


In addition to X-ray emission originating from point sources, diffuse X-ray emission has also been reported in the direction of a few GCs.
Extended diffuse X-rays from the core are in many cases likely to be composed of many faint unresolved point sources (as discussed for 47 Tuc, \citealt{Grindlay02}; NGC 6440 \citealt{Pooley02b}; M80 \citealt{Heinke03c}; and for 10 different clusters by \citealt{Hui09}).  

More interesting is diffuse X-ray emission that is spatially distinct from the known point sources, and thus unlikely to be produced by fainter point sources. 
For example, \citet{Okada2007} 
detected diffuse \chandra X-ray emission 
near (within a few arcminutes, but not centered on the cluster) 
six GCs, including 47 Tuc (see also \citealt{Krockenberger1995} for 
a {\it ROSAT} detection), NGC 6752, M5, Omega Centauri, M80, and NGC 6266. 
Several diffuse X-ray sources 
are known to be background sources. 
\citet{Okada2007} identify the extended X-ray source towards $\omega$ Cen as a cluster or group at redshift 0.08, based on the redshift of 0.08 from a thermal plasma fit with redshift free, and general consistency of the fitted parameters (e.g. temperature and radius) with expectations for a background galaxy cluster.
The diffuse X-ray source initially thought to be associated with 47 Tuc was subsequently identified as a background cluster of galaxies with a redshift of 0.34$\pm$0.02 \citep{Yuasa2009}. 
\citet{Heinke2020} identified extended X-ray emission projected onto NGC 6366 \citep{Bassa08} with a background cluster of galaxies. \citet{Cheng21} identified the two extended X-ray sources near NGC 6752 \citep{Okada2007} as galaxy clusters at $z=0.239$ and 0.375.  

\citet{Wu2014} 
detected diffuse X-ray emission within the half-light radius of 47 Tuc via \chandra observations, 
fitting the 
spectrum 
with a 
power-law (PL) model with a photon index of $\Gamma\sim$1.0 plus a plasma model with a plasma temperature $kT\sim0.2$ keV. They suggest that this X-ray emission may be produced by a shock between the stellar wind and the interstellar medium, and/or a pulsar wind from the MSPs in the cluster. 

\ter is 
the GC with perhaps the most interesting detection of diffuse isotropic X-ray emission. 
\citet{Eger2010} analyzed 31 ks of \chandra observations and found an excess of diffuse X-ray emission outside the half-light radius of \ter. 
The spectral analysis of this diffuse emission indicated a non-thermal component (after subtracting the 
diffuse Galactic background emission; 
\citealt{Ebisawa2005}), which can be fitted by a PL model with $\Gamma=0.9\pm0.5$ \citep{Eger2010}.
However, with limited statistics, the origin of this diffuse X-ray emission was 
unclear. 
Subsequently, \citet{Eger2012} 
searched for diffuse X-ray emission from the fields of another six GCs (M62, NGC 6388, NGC 6541, M28, M80, and NGC 6139) that were detected in $\gamma$-rays.
However, none of those six GCs exhibited a significant excess of diffuse X-rays.

While the nature of diffuse X-ray emission from GCs is still unclear, 
several scenarios have been proposed to interpret the origin of this emission.
For instance, the relative motion between a GC and the Galactic plane might produce a bow shock along the direction of motion of the cluster, generating soft, thermal X-ray emission \citep{Krockenberger1995,Okada2007}.
Non-thermal X-ray emission may be produced by highly energetic electrons through synchrotron radiation.
Also, inverse Compton emission can generate non-thermal X-rays in GCs by scattering visible photons from charged particles accelerated in shock fronts \citep{Bell1978,Blandford1987}, as proposed by \citet{Krockenberger1995}.
And the populations of MSPs residing in GCs are considered to continuously provide such seed electrons \citep{Bednarek2007}.

Several works have attempted to model a wind of relativistic electrons from a population of pulsars in globular clusters, fitting this data to the diffuse X-ray emission along with other multiwavelength constraints. Key to this modeling is the detection of Terzan 5 in the GeV range \citep{Kong10,Abdo10}, which is generally attributed to the sum of pulsed curvature radiation from electrons within the pulsar magnetospheres, though alternative, inverse Compton, interpretations have been suggested \citep{Cheng10}. 
TeV gamma-rays from near Terzan 5 have also been detected \citep{HESS2011}, and suggested to be related to a pulsar wind from the cluster \citep[see particularly][]{Bednarek14}. Finally, extended radio emission has been detected near Terzan 5, but given the complexity of the Galactic plane in radio it is not certain that the extended radio emission is associated with Terzan 5 \citep{Clapson11}. 

Detailed state-of-the-art modeling of the high-energy emission from particles accelerated by pulsars in Terzan 5, and fit to constraints on the diffuse X-ray, GeV, and TeV data, have been performed by \citet{Kopp13} and \citet{Ndiyavala19}. In these models, X-rays are produced by synchrotron radiation, GeV gamma-rays by curvature radiation, and TeV gamma-rays by inverse Compton radiation from the combined pulsar wind. \citet{Kopp13} were unable to replicate the spectral index of the diffuse X-ray emission from Terzan 5 as found by \citet{Eger2010}. \citet{Ndiyavala19} were able to approach the inferred X-ray spectral index, but at the cost of requiring a new {spectral energy density} (SED) component, high-energy synchrotron radiation (HESR) with a cutoff energy of 100 keV. Reaching these energies required \citet{Ndiyavala19} to place these electrons inside the magnetospheres of the Terzan 5 pulsars, to {increase} the magnetic field enough to reach the required maximum energy. The downside of this scenario is that hard X-rays from the population of Terzan 5 pulsars would be identified as bright point sources, unless they are exceptionally numerous. For instance, spreading hard X-rays across an area of 40 square arcminutes would require 144,000 pulsars if they are distributed evenly at one per square arcsecond. This makes the HESR scenario, as developed, very unlikely, and motivates us to doublecheck the diffuse X-ray data on Terzan 5.


Since the diffuse X-ray emission from \ter was first reported by \citet{Eger2010}, a number of \chandra observations of \ter have been performed (e.g. \citealt{Degenaar15}), motivating us to reanalyze and verify the existence of this emission.
In this work, we retrieve all the available and useful \chandra observations of \ter, and conduct a deep study of the diffuse X-ray emission from the field of \ter.
We also investigate the nature of the diffuse X-ray emission.
This paper is organized as follows.
In Section~2, we describe the process of \chandra data selection, preparation, and reduction.
We present primary spectral fitting results and analysis in Section~3.
Finally, we discuss the nature of the diffuse X-ray emission from \ter and draw conclusions in Section~4.

\section{Data preparation and reduction}


We prepared and reduced the \chandra datasets using {\sc ciao}\footnote{Chandra Interactive Analysis of Observations, available at \url{https://cxc.cfa.harvard.edu/ciao/}.} (version 4.15.2 with {\sc caldb} 4.10.7; \citealt{Fruscione2006}), the software package developed by the Chandra X-ray center (CXC).

\subsection{\chandra observations of \ter}

\ter has been observed by \chandra Advanced CCD Imaging Spectrometer (ACIS) in the timed exposure mode and FAINT telemetry format 24 times, with a total exposure time of 833.0 ks over a time span of more than twenty years (see Table~\ref{tab:observations}). 
Among these observations, 
six 
(ObsIDs 654, 655, 11051, 12454, 13708, and 27736) 
included an X-ray binary in a bright transient outburst \citep[e.g.][]{Heinke2003,Altamirano2012,Bahramian2014}, which severely affected the detectability of diffuse emission in the field of \ter. For our purpose in this work, therefore, these six observations were not included in the following analysis.
In addition,  ObsID 3798 suffered from strong background flares \citep[described in][]{Heinke2006}, leading to significantly increased background levels compared to quiescent background periods. However, this observation is eligible for our research purpose after removing the periods of flaring background (see Section~\ref{subsubsec:remove_flares}), and hence it was used in this work. 

Consequently, 18 observations were selected for further analysis. 
We obtained the datasets of these 18 observations from the Chandra Data Archive\footnote{\url{https://cxc.cfa.harvard.edu/cda/}}, and reprocessed them to generate new level=2 event files using the {\tt chandra\_repro} script\footnote{\url{https://cxc.cfa.harvard.edu/ciao/threads/createL2/}}. 
Note that the remaining observations were all observed using solely the ACIS-S3 chip, with other chips turned off, except for ObsID 10059, which used the S2 and S4 chips as optional CCDs. 
However, we restricted {our} attention to the S3 chip for observation 10059 to maintain consistency.
Observations 10059, 14475, 14476, 14477, 14478, and 14479 were performed in a subarray mode, where only partial regions of the S3 chip were used to take data.  

\begin{table}
    \centering
    \caption{Available \chandra observations of \ter}
    \resizebox{\columnwidth}{!}{%
    \begin{tabular}{lcccc}
    \hline
ObsID	&	Date	&	Exposure	& Filt. exp.$^a$ &	Instrument	\\
 & (yyyy-mm-dd) & (ks) & (ks) & \\
\hline
655*	&	2000-07-24	&	42.2	&	--	&	ACIS-I	\\
654*	&	2000-07-29	&	5.0	&	--	&	ACIS-I	\\
3798$^\dag$	&	2003-07-13	&	39.3	&	23.5	&	ACIS-S	\\
10059	&	2009-07-15	&	36.3	&	33.4	&	ACIS-S	\\
11051*	&	2010-10-24	&	9.9	&	--	&	ACIS-S	\\
13225	&	2011-02-07	&	29.7	&	28.4	&	ACIS-S	\\
13252	&	2011-04-29	&	39.5	&	37.0	&	ACIS-S	\\
13705	&	2011-09-05	&	13.9	&	12.9	&	ACIS-S	\\
14339	&	2011-09-08	&	34.1	&	33.8	&	ACIS-S	\\
12454*	&	2011-11-03	&	9.8	&	--	&	ACIS-S	\\
13706	&	2012-05-13	&	46.5	&	43.6	&	ACIS-S	\\
13708*	&	2012-07-30	&	9.8	&	--	&	ACIS-S	\\
14475	&	2012-09-17	&	30.5	&	18.7	&	ACIS-S	\\
14476	&	2012-10-28	&	28.6	&	21.2	&	ACIS-S	\\
14477	&	2013-02-05	&	28.6	&	18.2	&	ACIS-S	\\
14625	&	2013-02-22	&	49.2	&	46.9	&	ACIS-S	\\
15615	&	2013-02-23	&	84.2	&	77.5	&	ACIS-S	\\
14478	&	2013-07-16	&	28.6	&	17.7	&	ACIS-S	\\
14479	&	2014-07-15	&	28.6	&	19.2	&	ACIS-S	\\
16638	&	2014-07-17	&	71.6	&	64.2	&	ACIS-S	\\
15750	&	2014-07-20	&	23.0	&	20.9	&	ACIS-S	\\
17779	&	2016-07-13	&	68.9	&	64.2	&	ACIS-S	\\
18881	&	2016-07-15	&	64.7	&	60.1	&	ACIS-S	\\
27736*	&	2023-03-18	&	10.6	&	--	&	ACIS-S	\\
\hline
    \end{tabular}
    }
    \tablefoot{\tablefoottext{a}{Filtered exposure time by removing flaring periods. See Section~\ref{subsubsec:remove_flares}.} \\
    \tablefoottext{*}{Observations severely affected by X-ray transient outbursts, and therefore not used in this work.} \\
    \tablefoottext{\dag}{A strong flare occurred during this observation (ObsID 3798).} }
    \label{tab:observations}
\end{table}

\subsection{Point source exclusion}
\label{subsec:source_excusion}

Given that 
our 
interest in this work is the diffuse X-ray emission in the field of \ter, it is therefore necessary to remove emission from any point sources. 
Point source exclusion is also required for checking for background flares. 

There are several existing catalogues of point sources in \ter \citep[e.g.][]{Heinke2006,Bahramian2020}. Specifically, \citet{Heinke2006} used observation 3798 with a total good time of 35.3 ks and detected 50 and 77 X-ray sources within and outside the half-light radius {(\rh)} of \ter, respectively. However, the currently available observations of \ter are much deeper than the single  observation used in \citet{Heinke2006}, and hence one can expect more sources to be detected in the field of \ter with a longer exposure time.
On the other hand, \citet{Bahramian2020} used a compilation of  datasets similar to those used in this work, and detected 188 confident sources with a total exposure of 597.0 ks within a 1-arcmin-radius region. The region covered in their work, however, is too limited for our study, as we are interested in a broader area out to a few arcminutes from the center of \ter. 
Therefore, to sufficiently exclude point sources, we performed source detection in the field of \ter with the combined observations in this work.

We first ran the {\tt merge\_obs} script\footnote{\url{https://cxc.cfa.harvard.edu/ciao/threads/merge_all/}}, reprojecting and combining the 18 selected observations to create a merged event file, exposure maps, and exposure-corrected images. 
We then applied {\tt wavdetect}\footnote{\url{https://cxc.cfa.harvard.edu/ciao/threads/wavdetect/}}, a Mexican-Hat Wavelet source detection tool, with scales of [1, 1.4, 2, 4, 8] pixels and a significance threshold of 10$^{-6}$ (suggesting one false source per $10^{-6}$ searched pixels), to detect sources in the field and generate a corresponding source list. 
We found a total of 333 sources outside the half-light radius and in the field of view (FoV) of the combined image. 
To better account for the \chandra point spread function (PSF), we implemented the {\tt psf} module\footnote{\url{https://cxc.cfa.harvard.edu/ciao/ahelp/psf.html}} to estimate the size of the \chandra PSF at the off-axis positions of the detected sources for each observation separately. We found 
that PSF-corrected regions 
obtained with an 
encircled energy fraction 
of 95\% at 1.5 keV were sufficient to exclude the emission from those point sources.

\subsection{Background removal}

Instrumental background should be removed carefully to ensure {that} the final extracted products (e.g. spectra) contain (almost) purely the information about diffuse X-ray emission. 
There are four components to the ACIS background\footnote{See details on the ACIS background in the \chandra Proposers' Observatory Guide, \url{https://cxc.cfa.harvard.edu/proposer/POG/html/chap6.html\#tth\_sEc6.17}, Chapter~6; and in \citet{Hickox2006}.}: 
(1) background flares caused by charged particles, (2) cosmic X-ray background (CXB), principally from distant active galactic nuclei, (3) non-celestial X-ray background (NXB), and (4) read-out artifacts.
Note that read-out artifacts are only significant for strong X-ray sources, whereas the detected X-ray sources in this work are relatively faint, making read-out artifacts negligible.
Therefore, read-out artifacts were not taken into account in the following analysis. 

\subsubsection{Flaring background}
\label{subsubsec:remove_flares}

\begin{figure*}
    \centering
    \includegraphics[width=1.1\columnwidth, valign=c]{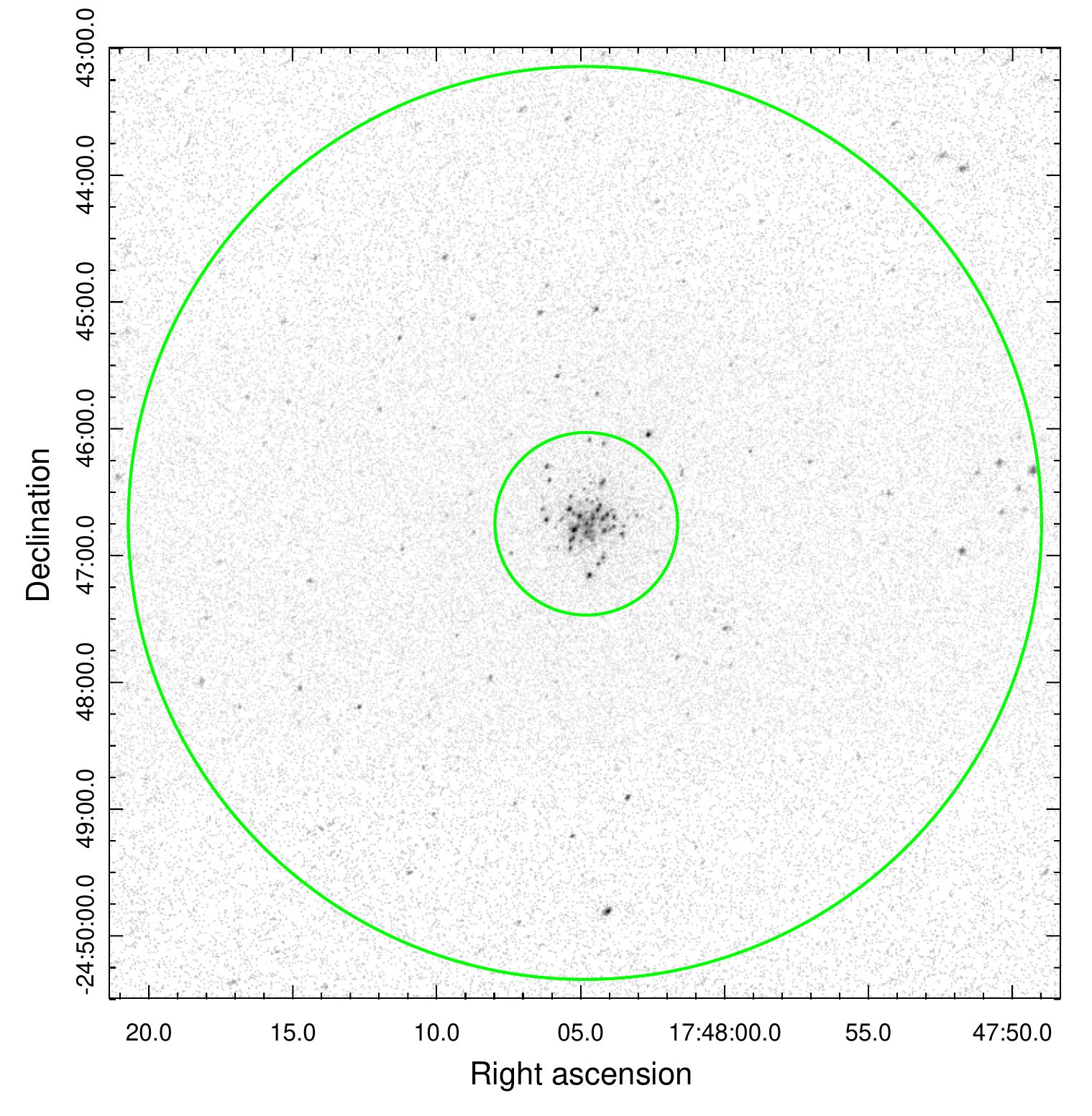}
    \includegraphics[width=0.9\columnwidth, valign=c]{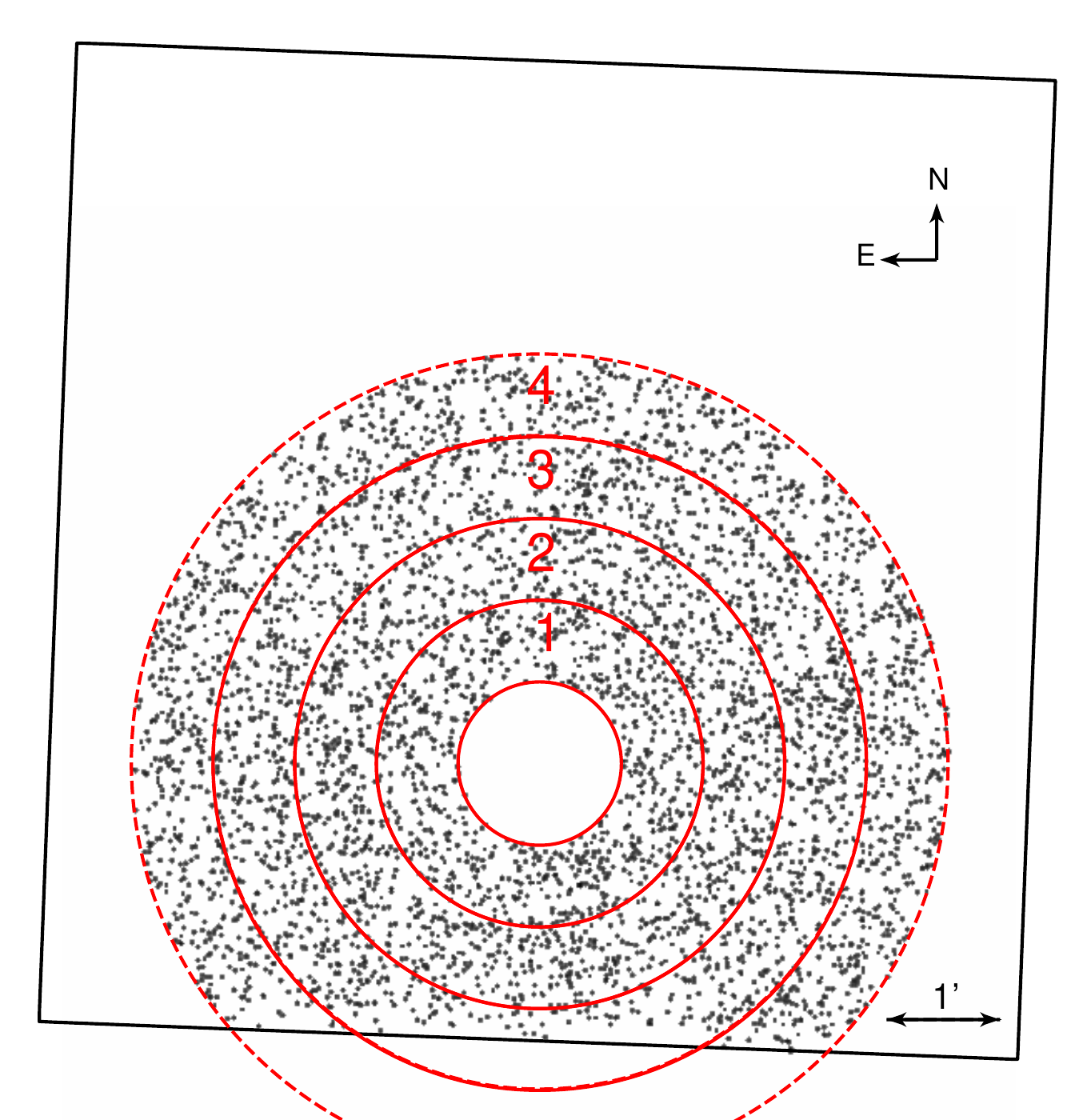}
    \caption{Left: Exposure-corrected, combined \chandra X-ray image of \ter in the band 0.5--7 keV. The small and large green circles show the half-light radius (0.72 arcmin) and five times the half-light radius (3.60 arcmin), respectively, centered at R.A.=17:48:04.80, Dec.=$-$24:46:45 (\citealt[2010 edition]{Harris1996}). Right: The region of interest (RoI) and the spectral extraction regions of the observation 3798 for an example. See the region definitions in Section~\ref{subsec:specextract}. The outer square shows the field of view (FoV) of the observation, while the regions of the {four} annuli within the FoV {(ring numbers annotated)} show the spectral extraction areas. {Ring 4 was used to extract \ter background and thus marked with a dashed circle.} The X-ray image is generated in the band 0.5--7 keV with {smoothed 0.492-arcsec pixels}. North is up, and east is to the left.}
    \label{fig:x-ray_image}
\end{figure*}

To remove {intervals with} background flares, we followed the {\tt lc\_clean} routine in a {\sc ciao} science thread\footnote{\url{https://cxc.cfa.harvard.edu/ciao/threads/flare/}}, which also removes periods of anomalously-low count rates.
In brief, for each observation, we extracted the light curve of the corresponding FoV, using {\tt dmextract}. We then applied {\tt lc\_clean} to select those quiescent regions of the light curve and create a new good time interval (GTI) file, which was used to filter the corresponding event file to get the flare-free observation.
In particular, since observation 3798 suffered from strong flares, for this case we forced the {\tt mean} option of {\tt lc\_clean} to be 0.37 counts s$^{-1}$ based on its count rate distribution. The filtered exposure time of each observation is listed in Table~\ref{tab:observations}, while the total filtered exposure time is 641.6~ks. 
Figure~\ref{fig:x-ray_image} (left panel) shows a combined X-ray image of \ter in {the energy band of} 0.5--7 keV using those filtered event files.

\subsubsection{CXB and NXB}
\label{subsubsec:blanksky}

The event files filtered with GTIs are essentially clean. 
However, the signals of CXB and NXB that are not relevant to the diffuse emission in \ter still remain in the event files. 
To determine and subtract the 
{NXB} for further spectral analysis, we used the {\tt blanksky} script to create blank-sky background files
{using observations in which the ACIS detector is stowed.\footnote{\url{https://cxc.harvard.edu/ciao/threads/acisbackground/}}
Each background file was scaled 
to match the 
background particle count rates in the band 9--12 keV, where the flux is almost entirely due to particle background \citep[see][]{Hickox2006}. The obtained blank-sky files were subsequently used to extract the spectra of the NXB. }

{We note that the {\tt blanksky} script is able to generate background files containing both NXB and CXB. However, those files are created using observations of relatively empty fields with the conditions of high galactic latitude (|b|>20$\degree$) and little soft X-ray emission, where soft ($\lesssim$2 keV) cosmic X-rays are barely absorbed. 
\ter, however, is located near the Galactic Center, and the hydrogen column density (\nh) towards it is about $2.0\times10^{22}~{\rm cm}^{-2}$.\footnote{This value of \nh was calculated based on the correlation between the optical extinction $A_V$ and \nh \citep{Bahramian2015}, while $A_V$ was estimated using the ratio $A_V/E(B-V)=3.1$ \citep{Cardelli1989}, where $E(B-V)=2.28$ is the foreground reddening towards \ter \citep[2010 edition]{Harris1996}. 
We note that $E(B-V)$ varies towards the field of \ter at different locations \citep{Massari2012}, but the value used here appears to reasonably match the reddening observed in the cluster regions we studied.}  
Thus, a large fraction of soft cosmic X-rays in the direction of \ter is extincted.
Using the background files containing both NXB and CXB to extract background spectra 
causes severe over-subtraction from the observations in the low-energy band.
Therefore, to properly subtract the CXB as well as other Galactic diffuse emission from the local environment around \ter, we defined the diffuse X-ray emission from the outer region of \ter (i.e. Ring 4; see below) as the local X-ray background (LXB), where the X-ray emission is assumed to be unassociated with \ter.  (This indicates that we may miss the most extended diffuse emission associated with Terzan 5; our results are a lower limit for this emission.) We then performed  spectral fitting to constrain its properties for subtraction purposes (see Section \ref{LXB_fit}).
}

\subsection{Spectral extraction}
\label{subsec:specextract}

To extract the X-ray spectra of the field of \ter from the observations,
we first defined the region of interest (RoI) to be an annular region from 1\rh to 5\rh (i.e. 0.72 to 3.60 arcmin) from the centre of \ter. 
{We chose these boundaries for two reasons:}
first, the diffuse X-ray emission in the region within the half-light radius can be 
largely 
explained as a result of unresolved sources due to the high source  density; and second, the region within 5\rh is fully contained in the combined FoV.
{Then we divided the RoI into four annuli with an equal width of 0.72 arcmin (or 1\rh; see the right panel of Figure~\ref{fig:x-ray_image} for an example), labeling the annuli as Ring 1 to Ring 4, progressing from the innermost to the outermost region.
We then extracted spectra from these rings separately for each observation with point source exclusion and FoV restriction, using the {\tt specextract} script for diffuse emission.\footnote{\url{https://cxc.cfa.harvard.edu/ciao/threads/extended/}} }
The background spectra were extracted from the blank-sky files using the same region definitions. 

Since the 
aim point and roll angle are different 
in each observation,
the FoVs and hence the actual areas used to extract spectra are also different. 
This effect could result in offsets in spectral intensities, 
especially for 
subarray-mode observations.
Therefore, to 
correct for 
the likely offsets, we calculated the actual area of each ring in each observation, and scaled each spectrum by the ratio of its actual area to the maximum actual area of the same ring(s) among all observations.
We list the calculated actual areas 
of each annulus in each 
observation in Table~\ref{tab:actual_areas}.

\section{Data analysis and results}

We performed the data analysis using {\sc sherpa} (version 4.15.0; \citealt{Freeman2001,Doe2007,Burke2022}), {\sc ciao}'s modelling and fitting package.

\subsection{The excess of diffuse X-ray emission}
\label{subsec:s2n}

\begin{figure}
    \centering
    \includegraphics[width=1.05\columnwidth]{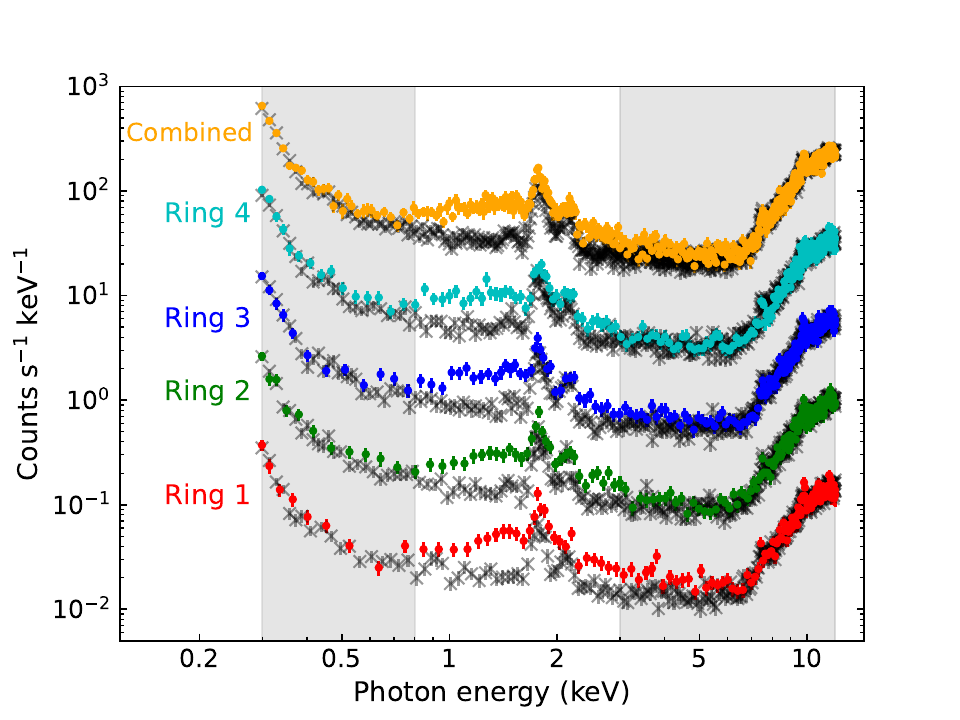}
    \caption{{The X-ray spectra of four rings and their combination from ObsID 15615 of the \ter field,  in the energy band 0.3--12 keV. The observed spectra (unsubtracted) are shown in colored dots, while the background spectra extracted from corresponding blank-sky files are shown in crosses, with photons grouped into 50 counts per bin. Shaded regions indicate where the observed and background spectra are almost indistinguishable. The spectra of each ring are scaled to provide adequate spacing for display purposes.}}
    \label{fig:obs_bkg_15615}
\end{figure}

To validate the background removal and verify the existence of X-ray excess in the field of \ter, we first 
{compared the observed, unsubtracted spectra with the corresponding blank-sky background spectra, for all the observations.
For each observation, we investigated the spectra extracted from the four rings separately, and also their combination.
}
{Each spectrum was grouped into 50 photons per bin in the 0.3--12 keV band. The spectra of ObsID 15651, 
which has the longest exposure time,
are shown as an example in Figure~\ref{fig:obs_bkg_15615}.} 

{The observed spectra are nearly indistinguishable 
from
the blank-sky background spectra in the bands $\lesssim 0.8$ keV and $\gtrsim 3$ keV, indicating
that 
the observed X-rays from these two bands are almost entirely contributed by the NXB. 
This also implies that the blank-sky files are well-scaled to match the corresponding observations. 
Two prominent emission line features, centered at around 1.8 keV for Si and 2.1 keV for Au, can be seen from the spectra, which are dominated by the NXB\footnote{See \url{https://cxc.cfa.harvard.edu/proposer/POG/html/chap6.html\#tth_sEc6.17.1}}. 
Nevertheless, significant X-ray excesses are observed in the band 0.8--3 keV in each ring. 
Therefore, in the following spectral analysis, we only focused on the band 0.8--3 keV to ensure sufficient statistics and significance. 
We note that, for the Ring 1 spectra, the X-ray excess might reach up to $\sim$5 keV (as shown in Figure~\ref{fig:obs_bkg_15615}).
However, for consistency and simplicity, we ignored the X-rays above 3 keV for all the spectral analysis.}

\subsection{Spectral analysis}
\label{subsec:spec_analysis}

Given that the 18 \chandra observations used in this work have a timespan of about 13 years, it is therefore inappropriate to create combined spectra for those observations for analysis, due to the accumulating contamination on the ACIS detectors and thus the changing corrections for this contamination.\footnote{\url{https://cxc.cfa.harvard.edu/ciao/why/acisqecontamN0015.html}}
Instead, we performed simultaneous fitting for these 18 datasets, allowing us to increase the significance of the fits without averaging spectra and responses.
We implemented $\chi^2$ statistics (with variance computed from data amplitudes) and a Monte-Carlo optimization method in the fitting process.
We considered  X-ray absorption by the interstellar medium towards \ter for all the spectral fits in this work, using the {\tt xstbabs} model {(TBABS)} with {\tt wilm} abundances \citep{Wilms2000} and {\tt vern} cross sections \citep{Verner1996}.

{\subsubsection{Local X-ray background (LXB) fitting}}
\label{LXB_fit}

\begin{figure}
    \centering
    \includegraphics[width=\linewidth]{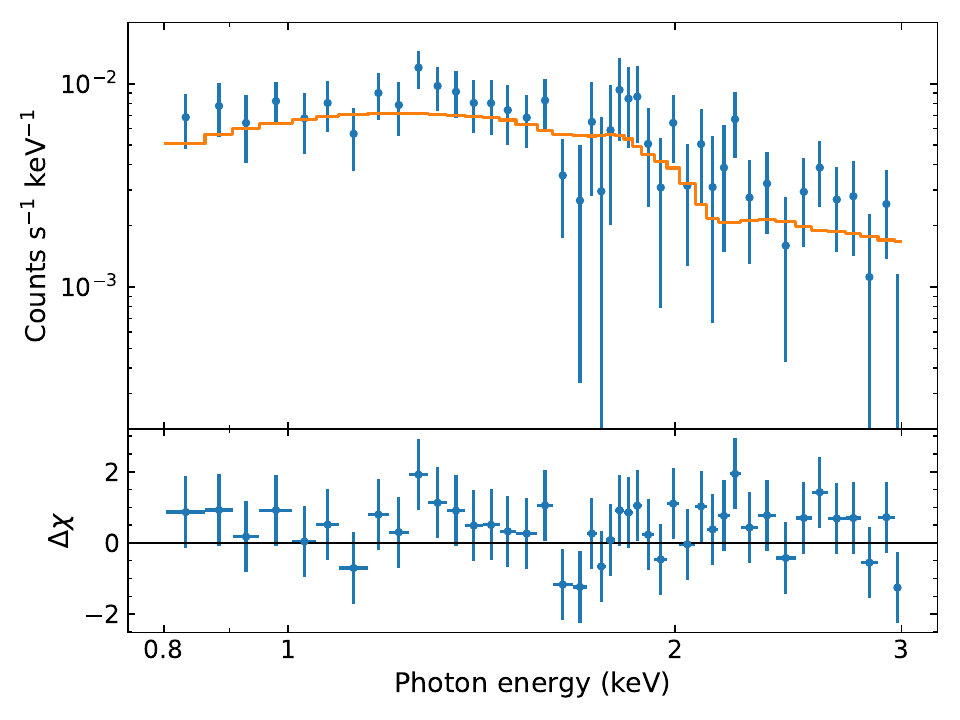}
    \caption{{The spectrum and the best fit of the diffuse X-ray emission (0.8--3 keV) extracted from Ring 4 of \ter from ObsID 15615 (see Section~\ref{subsec:specextract} for the ring definition). The best-fit model is given by a single absorbed power law. Data are grouped into at least 50 photons per bin.}}
    \label{fig:ring4_spec_15615}
\end{figure}

{As we aim to verify the diffuse X-ray emission that is truly associated with \ter, it is necessary to identify and exclude any local X-ray background (LXB), for instance from interstellar gas in our Galaxy along the line of sight.
Therefore, we first fitted the spectra extracted from Ring 4, where the X-ray emission is assumed to represent the LXB, with a simple absorbed PL model (TBABS$\times$PL), allowing the \nh value to vary. The fitting results are summarized in Table~\ref{tab:ring4_fits}. An example spectrum, and fit, of ObsID 15615 are shown in Figure~\ref{fig:ring4_spec_15615}.
}

{The Ring 4 spectra can be well described by a single PL model with a photon index $\Gamma = 1.44 \pm 0.20$. The indicated unabsorbed X-ray surface flux in the band 0.5--6 keV is $(3.96\pm0.18)\times10^{-18}~{\rm erg~cm^{-2}~s^{-1}~arcsec^{-2}}$. The best-fit \nh value is $1.4\times10^{21}~{\rm cm}^{-2}$, approximately an order of magnitude lower than the value towards \ter, implying that most, if not all, of the diffuse X-ray emission from Ring 4 originates from the foreground relative to Terzan~5. 

Intriguingly, the PL photon index of the LXB spectra is consistent with that of unresolved CXB spectra extracted from the Chandra Deep Field (CDFs) North and South observations, which are targeted toward regions of low \nh and away from bright features in the Galactic emission ($\Gamma=1.5^{+0.5}_{-0.4}$; \citealt{Hickox2006}).
The surface flux of the CXB from CDFs is $3.4\times10^{-19}~{\rm erg~cm^{-2}~s^{-1}~arcsec^{-2}}$ in the band 1--8 keV, about an order of magnitude lower than that of the LXB, which is expected since the CXB towards the Galactic Center area is much higher than the CXB toward the CDFs (see e.g. \citealt{Snowden1997}). 
The LXB likely has significant contributions from the local hot bubble, the Galactic halo and corona, unresolved Galactic sources, etc.
}

{We did not fit the LXB with a two-temperature non-equilibrium ionization collisional plasma (NEI) model, as done by \citet{Eger2010}, 
as we found that an absorbed PL model can already adequately describe the spectra. Given that the LXB component is only used to 
subtract the foreground 
X-rays 
from the 
inner rings of \ter, complex spectral fittings are unnecessary.
}

\begin{table}
    \centering
    \caption{{Spectral fits of Ring 4 spectra with an absorbed power-law model}}
    \begin{tabular}{ccccc}
    \hline
        No. Ring & \nh$^a$ & $\Gamma$ & ${F_{\rm surf}}^b$ & dof/$\chi_\nu^2$ \\
    \hline
        4 & $0.14\pm0.10$ & 1.44$\pm$0.20 & 3.96$\pm$0.18 & 314/0.99 \\
    \hline
    \end{tabular}
    \tablefoot{{Errors are quoted at 1 sigma. \\
    \tablefoottext{a}{Hydrogen column density in units of $10^{22}~{\rm cm}^{-2}$}. \\
    \tablefoottext{b}{Unabsorbed X-ray surface flux in the band 0.5--6 keV in units of $10^{-18}~{\rm erg~cm^{-2}~s^{-1}~arcsec^{-2}}$}. }}
    \label{tab:ring4_fits}
\end{table}

{\subsubsection{Diffuse X-ray emission associated with Terzan 5}}

{To identify the diffuse X-ray emission that is truly associated with \ter, we analyzed the spectra extracted from the inner three rings,  excluding  the LXB contribution.
However, instead of directly subtracting the LXB spectra from the inner rings' spectra, we included the LXB contribution as a fixed component with the best-fit parameters in the fits, while the \nh towards the LXB was still set to be a free parameter to allow for possible changes. 
{We fixed the normalization (or flux) of the LXB contribution to each ring based on the area of each ring, to give the same normalization per unit area of the LXB for each of the inner rings as seen in ring 4.}
For the potential diffuse X-ray component from \ter, we assumed the \nh towards \ter to be fixed at $2.0\times10^{22}~{\rm cm}^{-2}$.
}

\begin{figure}[h!]
    \centering
    \includegraphics[width=\columnwidth]{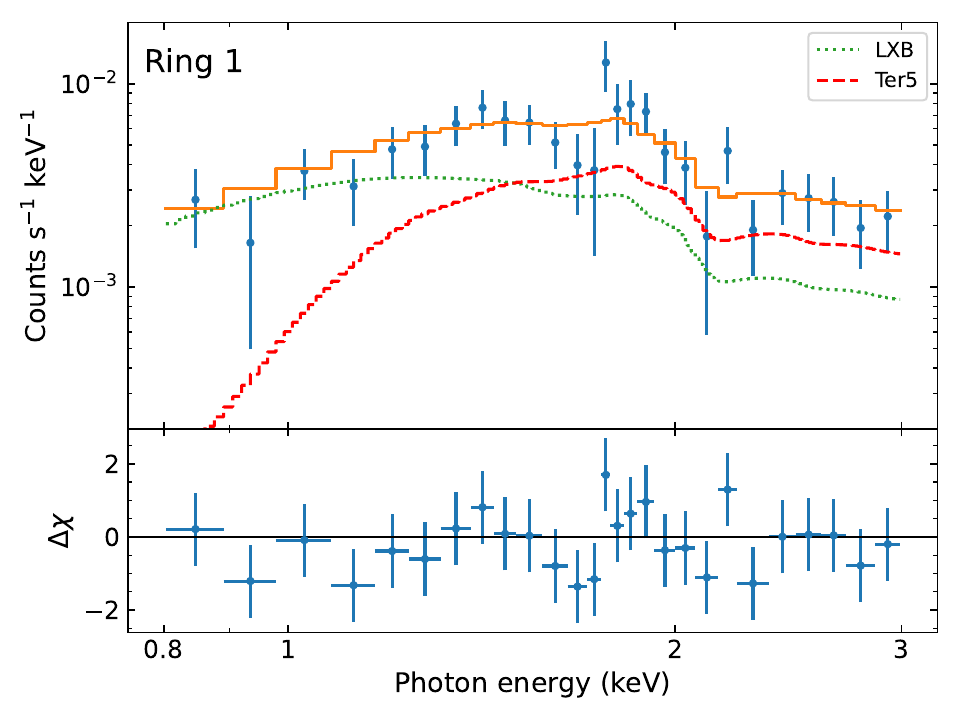}
    \includegraphics[width=\columnwidth]{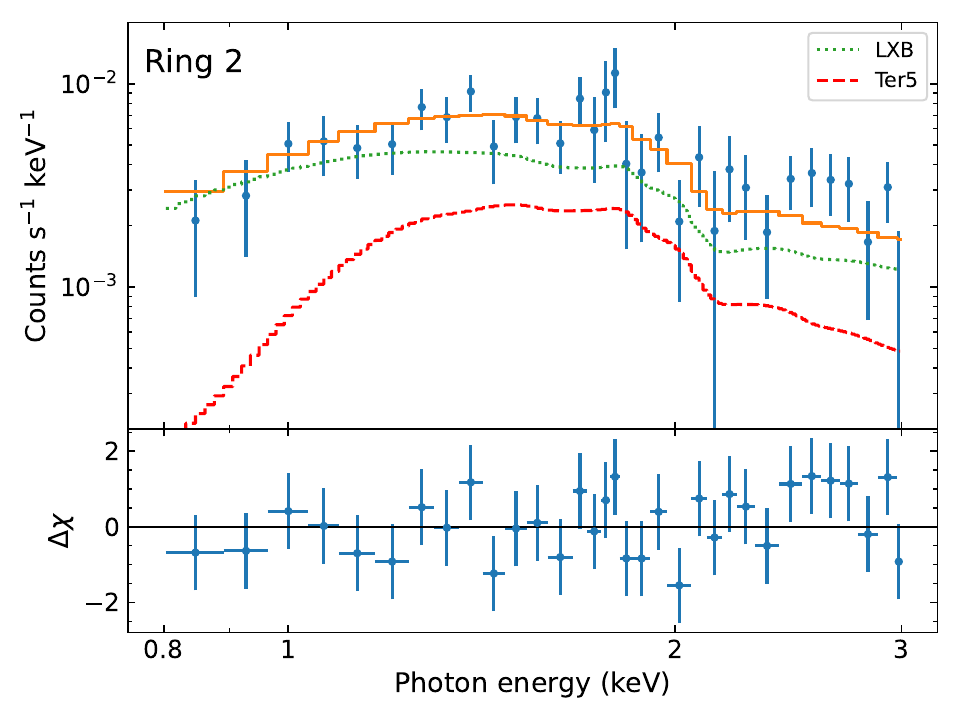}
    \includegraphics[width=\columnwidth]{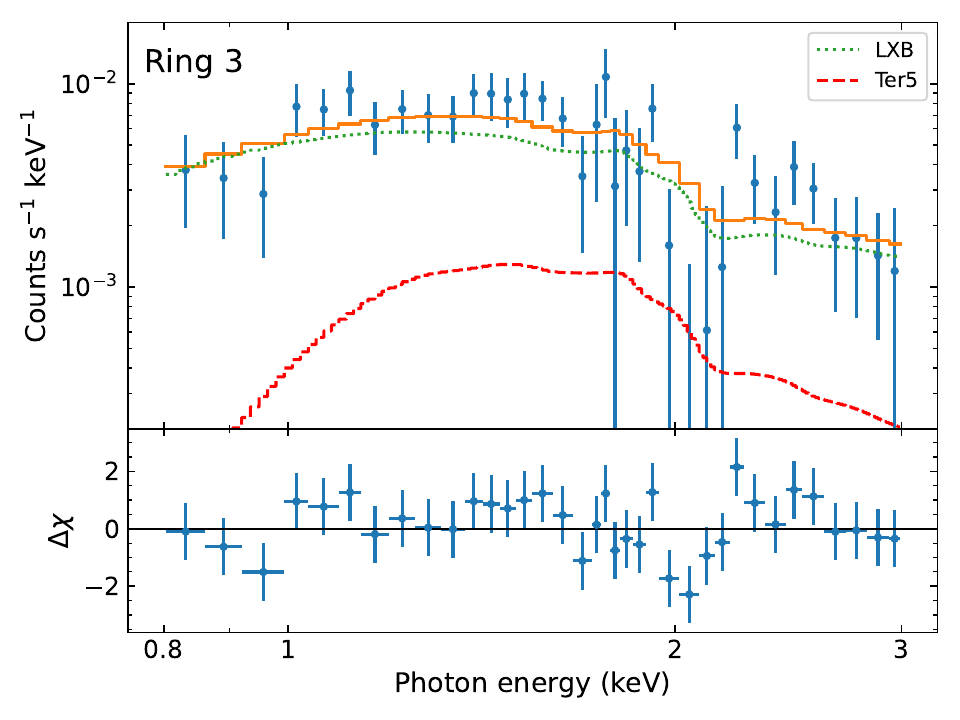}
    \caption{{The spectra and best fits of the diffuse X-ray emission (0.8--3 keV) extracted from inner three rings of \ter from ObsID 15615. The dotted line and the dashed line show the X-ray components of the local X-ray background (LXB) and \ter, fitted with absorbed power-law models with independent \nh values, respectively. The ring number is annotated at the upper-left corner. Data are grouped into at least 50 photons per bin.}}
    \label{fig:inner_spec}
\end{figure}

{We first fitted the spectra from the inner three rings with a double absorbed PL model: (TBABS$\times$PL)$_{\rm LXB}$+(TBABS$\times$PL)$_{\rm Ter5}$, where LXB and Ter5 denote the diffuse X-ray components from the LXB and \ter, respectively. 
The fitting results are listed in Table~\ref{tab:inner_fits}. We show the spectra and best fits of the inner three rings of ObsID 15615, as an example, in Figure~\ref{fig:inner_spec}.
}

\begin{figure}[th]
    \centering
    \includegraphics[width=\columnwidth]{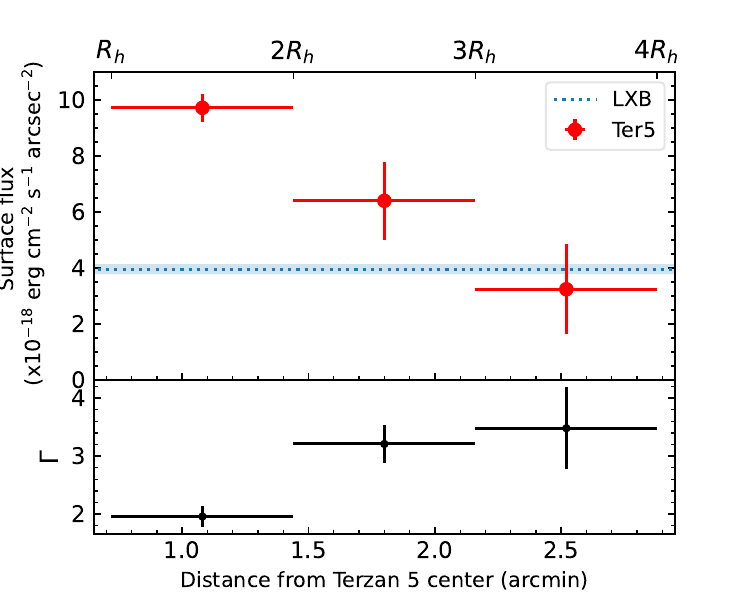}
    \caption{{The change of unabsorbed surface flux (upper panel) and the photon index (lower panel) of the diffuse X-ray emission from \ter, as a function of distance from the center of \ter. The horizontal error bars show the spatial extent of the rings, and the vertical error bars show 1$\sigma$ uncertainties. The unabsorbed surface flux of the local X-ray background (subtracted from the \ter5 flux measurements shown) is also shown in the upper panel (dotted line with 1$\sigma$ shaded region), which is assumed to be constant towards the \ter field.}} 
    \label{fig:SB-Gamma-Dist}
\end{figure}

{We found that, besides the diffuse X-ray emission from the LXB, significant X-ray emission is detected  associated with \ter in the inner three rings, and well fit by an absorbed PL model.
Moreover, the fitted photon indices increase from $\Gamma = 1.96 \pm 0.18$ (Ring 1) to $\Gamma = 3.48 \pm 0.71$ (Ring 3), indicating substantial steepening of the PL spectra from Ring 1 to Ring 3.
The unabsorbed X-ray surface flux of the \ter diffuse emission declines gradually from around $9.7 \times 10^{-18}~{\rm erg~cm^{-2}~s^{-1}~arcsec^{-2}}$ in Ring 1 to $3.2 \times 10^{-18}~{\rm erg~cm^{-2}~s^{-1}~arcsec^{-2}}$ in Ring 3, in the band 0.5--6 keV (see Figure~\ref{fig:SB-Gamma-Dist}).
On the other hand, there are no significant changes in the \nh values towards the LXB.
Additionally, we investigated the possible spatial variation of the diffuse emission by dividing Ring 1 into four quarters. No significant variation in the surface fluxes of those four quarters was found, indicating the diffuse emission is likely isotropic.
}

{The fitted PL photon indices of $\Gamma \gtrsim 2$ may also indicate a thermal origin of the observed X-rays. 
In the thermal scenario, electrons can lose energy and radiate X-ray photons through bremsstrahlung radiation.
To examine this possibility, we fitted the spectra by replacing the PL model with a bremsstrahlung model (BREM), i.e., (TBABS$\times$PL)$_{\rm LXB}$+(TBABS$\times$BREM)$_{\rm Ter5}$.
The fitting results are listed in the lower panel of Table~\ref{tab:inner_fits}.
}

\begin{table}
    \centering
    \caption{{Spectral fits of inner three rings}}
    \resizebox{\columnwidth}{!}{%
    \begin{tabular}{cccccc}
    \hline
     & \multicolumn{4}{c}{Model: (TBABS$\times$PL)$_{\rm LXB}$+(TBABS$\times$PL)$_{\rm Ter5}$} \\
        No. Ring & ${N_{\rm H, LXB}}^a$ & $\Gamma_{\rm Ter5}$ & ${F_{\rm surf, Ter5}}^b$ & dof/$\chi_\nu^2$ \\
    \hline
        1 & 0.27$\pm$0.05 & 1.96$\pm$0.18 & 9.73$\pm$0.50 & 240/1.02 \\
        2 & 0.33$\pm$0.05 & 3.21$\pm$0.33 & 6.41$\pm$1.40 & 285/1.07 \\
        3 & 0.22$\pm$0.04 & 3.48$\pm$0.71 & 3.24$\pm$1.63 & 284/1.07 \\
    \hline
    \hline
     & \multicolumn{5}{c}{Model: (TBABS$\times$PL)$_{\rm LXB}$+(TBABS$\times$BREM)$_{\rm Ter5}$} \\
        No. Ring & ${N_{\rm H, LXB}}^a$ & ${kT_{\rm Ter5}}^c$ & Norm$^d$ & ${F_{\rm surf, Ter5}}^b$ & dof/$\chi_\nu^2$ \\
    \hline
        1 & 0.26$\pm$0.04 & 3.11$\pm$0.83 & 4.84$\pm$0.63 & 8.29$\pm$1.62 & 240/1.02 \\
        2 & 0.31$\pm$0.04 & 0.99$\pm$0.19 & 9.89$\pm$3.16 & 4.07$\pm$1.68 & 285/1.06 \\
        3 & 0.21$\pm$0.04 & 0.86$\pm$0.31 & 6.94$\pm$4.71 & 1.70$\pm$1.33 & 284/1.06 \\
    \hline
    \end{tabular}
    }
    \tablefoot{{LXB and Ter5 denote the diffuse X-ray components from the local X-ray background and \ter, respectively. Errors are quoted at 1 sigma. \\
    \tablefoottext{a}{Hydrogen column density in units of $10^{22}~{\rm cm}^{-2}$}. \\
    \tablefoottext{b}{Unabsorbed X-ray surface flux in the band 0.5--6 keV in units of $10^{-18}~{\rm erg~cm^{-2}~s^{-1}~arcsec^{-2}}$}. \\
    \tablefoottext{c}{The plasma temperature in keV.} \\
    \tablefoottext{d}{The normalization of the BREM model in units of 10$^{-5}$ cm$^{-5}$, given by $3.02\times10^{-15} \int n_e n_I dV / 4 \pi D^2$, where $D$ is the distance to the source in cm and $n_e$, $n_I$ are the electron and ion densities in cm$^{-3}$.}}}
    \label{tab:inner_fits}
\end{table}

\section{Discussion}

\subsection{Comparison with Eger et al.}
{Through spectral analysis of the spectra extracted from four rings around the center of \ter, we successfully verified the existence of diffuse X-ray emission that is genuinely associated with \ter \citep{Eger2010}. 
Moreover, the \ter diffuse emission can be well described by an absorbed PL model, and more intriguingly, its spectra show significant steepening from the inner region to the outer region, with the best-fit photon indices increasing from around 2.0 to 3.5.
}

{
However, the PL photon indices of the \ter diffuse X-ray component found in this work are much softer than the one reported by \citet{Eger2010}, who found a PL photon index of $\Gamma = 0.9 \pm 0.5$ for the spectra of the diffuse emission between 0.83 arcmin and 3 arcmin in the 1--7 keV band.
The discrepancy of photon indices between our work and \citet{Eger2010} 
may be 
caused by our different spectral model assumptions.
\citet{Eger2010} applied a single, fixed \nh value of $1 \times 10^{22}~{\rm cm}^{-2}$ for the absorption of both their LXB component and \ter X-ray component, whereas we used separate \nh values towards the LXB (allowed to vary) and \ter X-ray component (fixed at $2 \times 10^{22}~{\rm cm}^{-2}$), respectively.
If we use similar spectral model settings with those in \citet{Eger2010}, the PL photon indices of the LXB and diffuse X-ray emission from Ring 1 are found to be $\Gamma = 2.92 \pm 0.12$ and $\Gamma = 0.39 \pm 0.14$, respectively, which are consistent with the results in \citet{Eger2010} within 1$\sigma$.
However, we argue that it may not be safe to fix the \nh value towards the LXB to be the same as the one towards \ter. 
First, the LXB likely contains diffuse X-rays from different origins and locations, and therefore, fixing the \nh value towards it may 
significantly alter 
the spectral fits. 
As mentioned by \citet{Eger2010}, their choice to fix the \nh value was a compromise due to the limited statistics in their spectra. 
However, in this work, we 
used 
much deeper \chandra observations towards \ter with significantly higher statistics, allowing us to easily treat \nh as a free parameter.
Indeed, the \nh values towards the LXB inferred from spectral fits range from $\sim1.4\times10^{21}~{\rm cm}^{-2}$ to $\sim3.3\times10^{21}~{\rm cm}^{-2}$, almost an order of magnitude lower than the \nh towards \ter. 
}

{\subsection{The origin of the diffuse X-ray emission from Terzan 5}}

{We found diffuse X-ray emission associated with \ter 
between 0.72 arcmin and 2.88 arcmin from \ter center, whose nature is of great interest. 
In this section, we briefly discuss several scenarios of the origin of this diffuse X-ray emission.
We adopt a distance to \ter from the Sun of 6.9~kpc \citep[2010 edition]{Harris1996} in the following. 
Hence the unabsorbed X-ray luminosity of the diffuse emission from the inner three rings is $(2.7\pm0.5)\times10^{33}$~\ergs in the 0.5--6 keV band.
}

{\subsubsection{Contribution from unresolved point sources}}
{
We first looked into the contribution from unresolved point sources from the \ter globular cluster. 
We adopted the spatial surface distribution of X-ray sources in \ter from \citet{Heinke2006}:
\begin{equation}
    S(r) = S_0 \left[1 + \left(\frac{r}{r_c}\right)^2 \right]^{(1-3q)/2},
\end{equation}
where $S_0$ is the normalization factor, $r$ is the distance from cluster center, $r_c$ is the core radius ($r_c = 0.16$~arcmin for \ter; \citealt[2010 edition]{Harris1996}), and $q$ is the ratio of the masses of the X-ray sources and the stars that contribute to the core radius ($q=1.43$ for \ter; \citealt{Heinke2006}).
Then we obtained that the X-ray luminosity contributed by unresolved point sources between {1\rh (0.72 arcmin) and 4\rh (2.88 arcmin)} is only about 1.4\% of the unresolved X-ray luminosity within \rh. 
\citet{Heinke2006} estimated that the X-ray luminosity from unresolved sources within \rh is approximately $2\times10^{33}$~\ergs, and therefore the X-ray luminosity contributed by unresolved sources between 0.72 arcmin and 2.88 arcmin is roughly $2.8\times10^{31}$~\ergs, two orders of magnitude lower than the measured X-ray luminosity {of the diffuse emission}.
Hence, the contribution from unresolved point sources can be neglected. 
}

{\subsubsection{Synchrotron radiation}}

{One possible origin of \ter diffuse X-ray emission is the non-thermal synchrotron radiation (SR) emission generated by relativistic elections, which radiate photons at the critical frequency $\nu_c\sim4.2\times10^{-9} \gamma^2 (B/1\mu{\rm G})$~GHz, where $\gamma$ is the Lorentz factor of electrons, and $B$ is the magnetic field strength.
Suppose the magnetic field strength is a few $\mu$G in the \ter field, then to generate the X-rays observed in the band of a few keV, electrons with Lorentz factors of $\sim10^{8}$ and hence energies of $E=\gamma m_e c^2 \sim 10^{14}$~eV, where $m_e$ is the electron mass and $c$ is the speed of light, are needed. 
However, high-energy electrons are depleted rapidly due to the SR cooling process, the timescale of which is given by the electron energy divided by the synchrotron energy-loss rate, $dE/dt \sim 6.6 \times 10^{-16} \gamma^2 (B/1\mu{\rm G})^2$~eV~s$^{-1}$.
Thus, the SR cooling timescale $\tau_{\rm SR}$ can be estimated by $\tau_{\rm SR} \sim 10^{13} \gamma^{-1} (1 \mu{\rm G}/B)^2$~yr, and for the few-keV X-rays in \ter, $\tau_{\rm SR} \sim 10^5$~yr.
Given that $\tau_{\rm SR}$ is much smaller than the age of \ter, continuous injection of such electrons is required to produce the observed diffuse emission.
The large number of MSPs residing within the \rh of \ter are highly likely to be the source of those relativistic electrons \citep[e.g.][]{Bednarek2007}.
On the other hand, the escape timescale of those electrons from the cluster center can be estimated using Bohm diffusion \citep{Venter2008} by $\tau_{\rm esc}=r_{\rm esc}^2/2\kappa_{\rm Bohm}$, where $r_{\rm esc}$ is the particle propagation distance, and $\kappa_{\rm Bohm}=c E/(3 q B)$, where $q$ is the particle charge.
Hence, we have $\tau_{\rm esc} \sim 10^{16} \gamma^{-1} (B/1 \mu{\rm G})(r_{\rm esc}/ 1 {\rm kpc})^2$~yr, and for the electrons from \ter center, the $\tau_{\rm esc}$ to 1\rh and 4\rh are about 200 yr and 3,000 yr, respectively.
The inferred $\tau_{\rm esc}$ are two to three orders of magnitude lower than the $\tau_{\rm SR}$, indicating that SR emission can indeed be produced in the field of \ter as observed.
}

{However, the interesting steepening in the spectra likely implies the increasing depletion of high-energy electrons from Ring 1 to Ring 3.
If the spectra steepening is indeed a result from SR cooling, like the ones observed from Crab Nebula \citep{Mori2004}, Mouse Nebula \citep{Klingler2018}, etc. (see \citealt{Reynolds2017} for a review), $\tau_{\rm SR} < \tau_{\rm esc}$ is needed. 
Hence, the magnetic field strength near \ter needs to be larger than a few $\mu$G. For example, for $B=10~\mu$G, $\tau_{\rm SR}$ and $\tau_{\rm esc}$ become $10^3$~yr and $2\times10^3$~yr (propagating to 1\rh), respectively, and therefore the electrons are cooled before escaping beyond 1\rh.
However, the SR cooling would eventually lead to a steepening in the electron spectral index of $\Delta p = 1$ \citep{Pacholczyk1970}, which translates to the change of photon index of $\Delta \Gamma = 0.5$. 
The observed change of photon indices from Ring 1 to Ring 3, however, is $\Delta \Gamma = 1.52\pm0.73$, larger than the expected value (within 2$\sigma$ though).
Although such a large change in photon index has also been observed from the Mouse Nebula ($\Delta \Gamma \approx 1.4$; \citealt{Klingler2018}), the proposed rapid SR cooling regime to produce this large variation needs an equipartition magnetic field $B \sim 200~\mu$G in the nebula, which 
seems to high to be plausible in a globular cluster 
(e.g. the GC 47 Tuc is found to have a magnetic field of $B\sim1 ~\mu$G; \citealt{Abbate2023}). 
Therefore, SR emission is unlikely to be the sole origin of the observed diffuse X-ray emission.
}

{\subsubsection{Thermal bremsstrahlung radiation}}

{
The Terzan~5 diffuse emission can also be well described by a thermal bremsstrahlung model, with the plasma temperatures and surface fluxes dropping significantly from Ring 1 to Rings 2 and 3.
Moreover, from the obtained normalizations of the BREM model, we can estimate the electron/ion number densities. 
We supposed that $n_e$ and $n_I$ are constant with $n_e \approx n_I$, and that the emission volume $V$ is a sphere with the radius  given by the size of the ring.
Hence, the estimated $n_e$ values are 0.19$\pm$0.07~cm$^{-3}$, 0.16$\pm$0.09~cm$^{-3}$, and 0.10$\pm$0.08~cm$^{-3}$, for Ring 1, Ring 2, and Ring 3, respectively.
Interestingly, these estimated values for Terzan~5 are  consistent with the $n_e$ value determined for the GC 47~Tuc \citep{Abbate2018}, making the thermal bremsstrahlung radiation  physically plausible as the origin of the \ter diffuse emission.
Additionally, the cooling timescale for the thermal bremsstrahlung radiation ($\tau_{\rm BREM}$) can be estimated by $\tau_{\rm BREM} = \varepsilon_{\rm tot} / \varepsilon^{ff}$, where $\varepsilon_{\rm tot} = 3(n_e + n_I) k T /2$ is the total kinetic energy density of the plasma, and $\varepsilon^{ff} = 1.4\times10^{-27} T^{0.5} n_e n_I Z^2 \Bar{g}_B$ is the total power per unit volume emitted by thermal bremsstrahlung in units of erg~s$^{-1}$~cm$^{-3}$, where $Z$ is the atomic number ($Z=1$ for hydrogen), and $\Bar{g}_B=[1.1, 1.5]$ is a frequency average of the velocity averaged Gaunt factor \citep{Rybicki1986}.
We therefore obtained $\tau_{\rm BREM} \sim 10^4 T^{0.5} / (n_e \Bar{g}_B)$ yr, and using the fitted values, we found $\tau_{\rm BREM} \sim 10^8$~yr for all the three rings. 
}

{Other possible origins of the \ter diffuse emission include inverse Compton emission, shocks, etc.
It is also likely that the diffuse emission may have contributions from multiple origins.
Follow-up studies of the nature of the \ter diffuse X-ray emission based on our results are highly encouraged.
}

{\section{Conclusions}}

In this study, we investigated the diffuse X-ray emission from the field of \ter using very deep \chandra observations, including 18 observations with a total exposure time of 641.6 ks.
We extracted X-ray spectra of the diffuse emission from {four} annular regions between 0.72~arcmin and 3.60~arcmin centered at R.A.=17:48:04.80, Dec.=$-$24:46:45 (\citealt[2010 edition]{Harris1996}), {with an equal width of 0.72 arcmin, which were labeled as Ring 1 to Ring 4, progressing from the innermost to the outermost region.}
{By comparing the extracted spectra with their corresponding background spectra generated by the {\tt blanksky} script, we found significant diffuse X-ray excess in the energy band 0.8--3 keV.}

{To identify the nature of the diffuse X-ray emission, we first defined the X-rays from Ring 4 as the local X-ray background, which can be well described by a power-law model with a photon index $\Gamma = 1.44 \pm 0.20$.
Furthermore, after removing the contribution from the local X-ray background, we found diffuse X-ray components that are truly associated with \ter from the inner three rings.
The spectral analysis shows that the \ter diffuse emission can be well fitted by a power-law model, and more intriguingly, the fitted photon indices increase significantly from $\Gamma = 1.96 \pm 0.18$ in Ring 1 to $\Gamma = 3.48 \pm 0.71$ in Ring 3.
This large variation, $\Delta \Gamma = 1.52 \pm 0.73$, indicates a substantial softening of the diffuse X-rays from Terzan 5 towards the outer regions.
We suggest that synchrotron radiation is a possible origin of the observed \ter diffuse emission, though the large steepening in the spectra is unlikely to be produced by solely synchrotron radiation cooling.
}

{We also found that the origin of the diffuse emission could be thermal bremsstrahlung radiation, with plasma temperatures of $kT=3.11\pm0.83$~keV in Ring 1 and of $kT\sim1$~keV in Rings 2 and 3. 
In addition, the gas densities implied by the thermal bremsstrahlung model are $\sim$0.1--0.2~cm$^{-3}$, consistent with the gas density measured from 47~Tuc \citep{Abbate2018}. 
Therefore, thermal bremsstrahlung is a possible contribution to the observed diffuse X-rays from \ter.
It seems likely that the diffuse X-ray emission may be a result of a mix of multiple radiation processes.
}

We note that the softer photon index of the diffuse non-thermal X-ray component we found is likely to be more consistent with a low-energy synchrotron radiation component produced by electrons in an extended wind from the Terzan 5 pulsars, as suggested by \citet{Kopp13}, rather than requiring a separate high-energy synchrotron radiation component from inside the pulsar magnetospheres, as postulated by \citet{Ndiyavala19}. 
{Future modelings and investigations of the \ter diffuse emission based on the results in this paper will prompt our understanding of its nature.}

\begin{acknowledgements}
{We thank the referee's useful and constructive comments.}
JZ is supported by China Scholarship Council (CSC), File No. 202108180023. 
CH is supported by NSERC Discovery Grant RGPIN-2023-04264. 
This research has made use of data obtained from the Chandra Data Archive, and software provided by the Chandra X-ray Center (CXC) in the application packages {\sc ciao}, {\sc sherpa}, {\sc ds9}, and {\sc pimms}.
This research has made use of NASA's Astrophysics Data System.
\end{acknowledgements}

%
\bibliographystyle{bibtex/aa.bst} 
\bibliography{ref} 
%

\begin{appendix}
\onecolumn
    \section{Actual areas of spectral extractions}

In Table~\ref{tab:actual_areas}, we list the calculations of the actual areas of spectral extraction regions for the 18 \chandra observations of \ter used in this work.

\begin{table*}[h!]
    \centering
    \caption{The actual areas of different regions of 18 \chandra observations of \ter.}
    \label{tab:actual_areas}
    \begin{tabular}{cccccc}
        \hline
        Obs. ID & \multicolumn{5}{c}{Actual areas (${\rm arcmin}^2$)} \\
         & Ring 1 & Ring 2 & Ring 3 & Ring 4 & Total \\
        \hline
3798	&	4.71	&	7.96	&	10.25	&	11.03	&	33.95	\\
10059	&	4.71	&	7.75	&	5.59	&	3.49	&	21.54	\\
13225	&	4.71	&	7.53	&	8.50	&	9.97	&	30.71	\\
13252	&	4.71	&	7.35	&	8.33	&	9.83	&	30.22	\\
13705	&	4.71	&	7.53	&	8.51	&	9.93	&	30.69	\\
14339	&	4.71	&	7.58	&	8.56	&	9.97	&	30.82	\\
13706	&	4.71	&	7.50	&	8.47	&	9.94	&	30.62	\\
14475	&	4.05	&	2.70	&	1.74	&	1.68	&	10.17	\\
14476	&	4.04	&	2.91	&	1.73	&	1.67	&	10.34	\\
14477	&	4.08	&	3.29	&	1.78	&	1.72	&	10.88	\\
14625	&	4.71	&	7.02	&	8.07	&	9.61	&	29.41	\\
15615	&	4.71	&	6.82	&	7.94	&	9.50	&	28.96	\\
14478	&	4.10	&	2.72	&	1.75	&	1.71	&	10.27	\\
14479	&	4.13	&	2.65	&	1.76	&	1.71	&	10.26	\\
16638	&	4.71	&	7.53	&	8.51	&	9.94	&	30.69	\\
15750	&	4.71	&	7.06	&	8.12	&	9.61	&	29.51	\\
17779	&	4.71	&	7.75	&	8.75	&	10.12	&	31.34	\\
18881	&	4.71	&	7.73	&	8.73	&	10.10	&	31.28	\\
\hline
\end{tabular}
\tablefoot{{The actual area is defined as the area used to extract the spectrum,  constrained in the corresponding field of view and with point sources removed. Rings 1--4 are the annular regions centered at R.A.=17:48:04.80, Dec.=$-$24:46:45 (\citealt[2010 edition]{Harris1996}), from 0.72 arcmin to 3.60 arcmin with an equal width of 0.72 arcmin. }}
\end{table*}

\FloatBarrier
\twocolumn

\end{appendix}

\end{document}